# Optical tweezers as a mathematically driven spatio-temporal potential generator

JOHN A.C. ALBAY,[1,4] GOVIND PANERU,[2,4] HYUK KYU PAK,[2,3] AND YONGGUN JUN [1,*]

[1] *Department of Physics, National Central University, Taoyuan, 320, Republic of China*
[2] *Center for Soft and Living Matter, Institute for Basic Science (IBS), Ulsan 44919, Republic of Korea*
[3] *Department of Physics, Ulsan National Institute of Science and Technology (UNIST), Ulsan 44919, Republic of Korea*
[4] *Co-first authors with equal contribution*
\* *yonggun@phy.ncu.edu.tw*

**Abstract:** The ability to create and manipulate the spatio-temporal potentials is essential in the diverse fields of science and technology. Here, we introduce an optical feedback trap system based on a high precision position detection and an ultrafast feedback control of a Brownian particle in the optical tweezers to generate spatio-temporal virtual potentials of the desired shape in a controlled manner. As an application, we study nonequilibrium fluctuation dynamics of the particle in a time-varying virtual harmonic potential and validate the Crooks fluctuation theorem in highly nonequilibrium condition.

## 1. Introduction

For the past three decades, there has been significant progress in the field of stochastic and information thermodynamics, where general laws such as fluctuation theorems and Jarzynski relations applicable to nonequilibrium phenomena have been discovered [1–3]. Many of these nonequilibrium relations are validated experimentally, thanks to the development of new technologies, which facilitates trapping and manipulation of Brownian particles, such as optical tweezers (OT). The OT has been a powerful tool for trapping and controlling Brownian particles in fluid [4,5]. It can trap and locate an object with subnanometer resolution and is capable of probing piconewton forces. As a result, it has been successfully used as an experimental tool in the diverse fields of science and engineering [6,7]. It has been used in biophysical experiments for the purpose of the position and force spectroscopy [8,9]. The OT has also been used in the field of nonequilibrium and information thermodynamics to demonstrate the validity of various fundamental relations [10–13]. For example, varying laser intensity with controlled artificial thermal noise allows one to demonstrate the Brownian nano-heat engine [10]. Placing two traps close enough can create a double-well potential to study Kramers' transition rate [12], stochastic resonance [11], and Landauer's principle of information erasure [14]. However, these prior studies could not modulate the barrier height and the tilt of the double-well potential in a controlled manner. Hence, despite the partial success of the optical tweezers in the study of the stochastic and information thermodynamics, its application is still limited when the generation of the mathematically-driven time-varying arbitrary shaped potential is required.

Recently, Cohen *et al.* developed a feedback-based technique called anti-Brownian electrokinetic (ABEL) trap by applying the feedback force in the form of electrophoretic force, which enables trapping of a nano-sized object in solution [15,16]. The ABEL trap can also create the arbitrarily-shaped potential [17,18] and has been used to study the dynamics of a Brownian particle in a double-well potential [19–21]. However, the design and the implementation of the ABEL trap are quite complicated. In particular, ABEL trap requires micro-fabricated 2D flow channel that introduces complicated boundary effect between the trapped particle and the wall of the flow channel. Also, due to the long delay time in feedback [21], the ABEL trap cannot apply ultrafast feedback control, thereby limiting its application in studying stochastic dynamics in highly nonequilibrium regimes.

In this article, we propose a simple and effective technique called the optical feedback trap (OFT) that, with a high precision position detection and ultrafast feedback control, can create a time-dependent mathematically-driven effective potential of ideally any desired shape and strength using optical feedback force. The basic operation of the OFT follows three crucial steps. First, a high precision measurement of the particle position in the optical tweezers is acquired. Second, the feedback force necessary for the generation of the virtual potential of the desired shape is computed. The feedback force is finally applied to the particle in the form of optical force via the ultrafast modulation of the trap center. We tested the creation of the virtual harmonic potentials of various stiffnesses at a fixed laser intensity and found that the dynamics of the particle in the virtual potential is very close to the real harmonic potential. We also created a virtual double-well potential whose barrier height and well depths can be modulated in a controlled manner. As an application of the OFT, we studied the nonequilibrium fluctuation dynamics of a particle in virtual harmonic potential where the stiffness of the potential was varied linearly with time (keeping the laser power fixed) to test the validity of the Crooks fluctuation theorem [22] for diverse processes ranging from near equilibrium to very far from equilibrium.

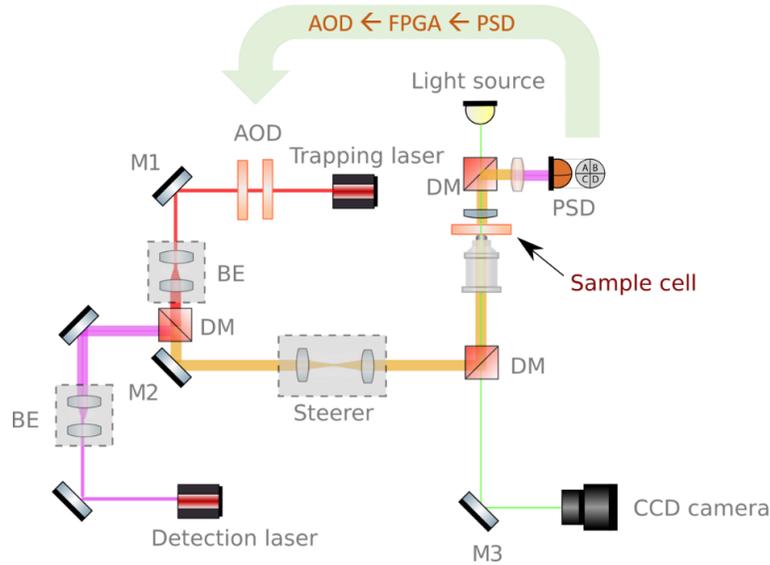

Fig. 1 Schematic drawing of the optical feedback trap. The particle position that the PSD acquires is sent to the FPGA which calculates the force exerting on a particle for the given virtual potential, convert it to the voltage, and send it to the AOD to deflect the beam. BE: beam expander, DM: dichroic mirror, PSD: position sensitive device, M: mirror, AOD: acoustoptic deflector, FPGA: Field-programmable gate array.

## 2. Experimental setup

The schematic drawing of the OFT setup is shown in Fig. 1. A laser (Cobolt Rumba) with 1064 nm wavelength is used for trapping the particle. The laser beam is incident on the acousto-optic deflector (AOD) (Gooch and Housego, AODF 4090-6) at Bragg angle, resulting in the maximum power output of the first-order diffracted beam. This beam is focused at the sample plane of an optical microscope (Olympus IX73) using a 100X oil immersion objective lens (Olympus, UPLFN100XO). A second laser (Thorlabs, BL976-SAG300) with 980 nm wavelength in combination with a neutral density filter (Thorlabs, NDUV10B) is used for detection of the particle position. The particle position is detected by a method based on the back-focal-plane interferometry [23]. Here, a condenser lens of high numerical aperture (NA

1.4) collects both scattered and unscattered detection laser light from the trapped particle and forms an interference pattern at the back focal plane of the condenser lens. The conjugate image of this pattern is projected onto the position-sensitive diode (PSD) (Pacific Silicon Sensor, DL100-7-PCBA3). The voltage signal acquired from the PSD is sent through a field programming gate array (FPGA) data acquisition board (National Instruments, NI PCIe-7851R) to convert the voltage to the real position of the particle and to determine the associated feedback force imposed by the virtual potential. The accuracy of the position measurement is about 1 nm. The FPGA board updates the tuning voltage that is needed for a shift of the laser beam center corresponding to above-determined feedback force. This voltage is applied to the AOD via the radio-frequency (RF) synthesizer driver (Gooch and Housego, AODR 1110FM-4) to steer the laser beam center. The particle position detection and application of the feedback are controlled via home-made software using LabVIEW programmed on the FPGA target. The trap stiffness is calibrated by two methods based on equipartition theorem and the power spectrum [24].

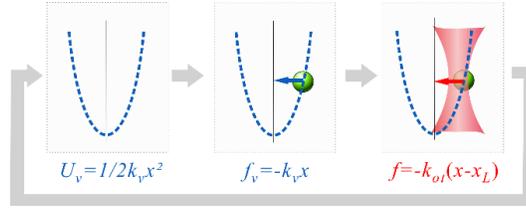

Fig. 2 Illustration of the protocol of the generation of a virtual harmonic potential. (a) A particle is assumed to be trapped in a virtual harmonic potential $U_v(x) = 1/2 k_v x^2$ (blue dashed curve) of stiffness $k_v$. (b) The PSD measures the particle position and the corresponding virtual restoring force $f_v = -k_v x$ (blue arrow) is estimated corresponding to the virtual potential. (c) The OT (red arrow) exerts the real feedback force $f = -k_{ot}(x - x_L)$ to the particle by shifting the potential center by the amount given by Eq. (1). Consequently, the particle moves as if it is in the real harmonic potential of stiffness $k_v$.

The basic operation for the generation of a virtual harmonic potential of the desired stiffness is outlined in Fig. 2. Initially, one can assume that a particle is trapped in the virtual harmonic potential centered at $x = 0$, $U_v(x) = 1/2\, k_v x^2$, experiences the virtual restoring force $f_v = -k_v x$, where $k_v$ is the stiffness of the virtual potential. Note that there is no real spatial potential. In order for the particle to feel itself in the potential, the optical tweezers exert the equal strength of physical force $f(t) = -k_{ot}[x(t) - x_L(t - t_d)]$ to the particle by shifting the laser center position $x_L$ instantaneously by an amount

$$x_L(t) = \left(1 - \frac{k_v}{k_{ot}}\right) x(t - t_d) = -\alpha x(t - t_d), \tag{1}$$

where $t_d$ is the feedback force delay time, which is defined as the time difference between the position measurement $x(t)$ and the actual application of the feedback force $f(t)$, and $\alpha \equiv -(1 - k_v/k_{ot})$ is the feedback gain. Consequently, the particle moves as if it is in the real harmonic potential of the stiffness $k_v$. The motion of the particle in this feedback trap can be described by the modified overdamped Langevin equation

$$\gamma \dot{x}(t) + k_{ot}\left[x(t) - x_L(t)\right] = \xi^f(t). \tag{2}$$

Here, $\gamma$ is the Stokes drag coefficient and $\xi^f$ is the random thermal Gaussian force with $\langle \xi^f \rangle = 0$ and $\langle \xi^f(t)\xi^f(t') \rangle = 2\gamma k_B T \delta(t - t')$, where $\delta(t - t')$ is the Dirac delta function.

The experimental procedure of the real-time feedback control system for realization of the virtual harmonic potential can be achieved as follows. The FPGA board generates the initial tuning voltage that locates the trap center to the initial position ($x_L(0) = 0$ for simplicity). This tuning voltage is applied to the AOD via RF synthesizer driver resulting in an OT at the sample plane of the microscope. The PSD measures the position of an optically trapped 1 $\mu$m diameter polystyrene particle. The voltage signal from the PSD is acquired to the FPGA, which computes the feedback force $f(t)$ corresponding to the shift of $x_L$ by the amount given by Eq. (1). The updated tuning voltage is applied to the AOD, the trap center is shifted to $x_L(t)$ in feedback updating time $t_u = 10$ µs, and another measurement-feedback-cycle is repeated.

Fig. 3(a) shows the trajectories of the particle diffusing in the virtual harmonic potential for 12 different values of $\alpha$, ranging from -0.77 (red) to 5.74 (purple) recorded at the constant stiffness of the OT, $k_{ot} = 41$ pN/µm. The negative (positive) value of $\alpha$ means that stiffness $k_v$ of the virtual harmonic potential is smaller (greater) than the stiffness $k_{ot}$ of the OT, and $\alpha = 0$ (green) corresponds to no feedback, i.e. $k_v = k_{ot}$. Fig. 3(b) shows the probability distributions of the particle position in virtual harmonic potential obtained from six different particle trajectories (pointed by black arrows) in Fig. 3(a). The solid curves are obtained by fitting the experimental data with the Boltzmann distribution whose variance is given by $\langle x^2 \rangle = k_B T / k_{\text{eff}}$, where $k_{\text{eff}}$ is the effective stiffness of the virtual harmonic potential. The corresponding virtual harmonic potentials depicted in Fig. 3(c) fit well (see solid curves) with the effective harmonic potential $U_{\text{eff}} = 1/2 k_{\text{eff}} x^2$. This shows that the virtual potentials generated by the OFT are well described with the mathematical harmonic potential.

Another quantity of interest is the measurement of $t_d$, which can be obtained from the power spectrum analysis of the trajectories in Fig. 3(a). The power spectrum density function $PS(f)$ of the feedback trap can be obtained from the Fourier transform of Eq. (2):

$$PS(f) = \frac{D}{\pi^2 \left| jf + f_c + f_c \alpha e^{-j2\pi f t_d} \right|^2}, \quad (3)$$

where $D = k_B T / \gamma$ is the diffusion coefficient, $j$ is the imaginary unit, and $f_c = k_{ot}/2\pi\gamma$ is the corner frequency. Note that for $\alpha = 0$, Eq. (3) reduces to the power spectrum density of the normal optical tweezers. The solid curves in Fig. 3(d) show the experimentally measured $PS(f)$ for six different values of $\alpha$ (pointed by black arrows in Fig. 3(a)). These measured data fit well with theoretical predictions (dashed curves) in Eq. (3), where $D = 1.1 \times 10^{-12}$ m$^2$/s and $f_c$= 1792 Hz obtained from the fitting of $PS(f)$ for $\alpha = 0$ were used. We obtained the feedback force delay time $t_d$= 20 µs for all $\alpha$, which corresponds to $2t_u$. For $-1 < \alpha \leq 1.5$, the $PS(f)$ follow the Lorentzian function. For $\alpha > 1.5$, $PS(f)$ deviates from the Lorentzian systematically until the particle overshoots the set point, leading to the damped oscillations, and a resonance in the power spectrum is observed for $\alpha > 6$. Even with the resonance peak, the corresponding Lorentzian spectra agree with our theoretical prediction.

To find the relation between the variance of position fluctuations and the feedback gain, we rewrite Eq. (2) regarding discrete position $x_n$ after the nth time step $nt_u$ as

$$x_{n+1} = x_n - \frac{2\pi k_{ot}}{\gamma} \left[ x_n - \left(1 - \frac{k_v}{k_{ot}}\right) x_{n-i} \right] t_u + \xi_n^x t_u. \quad (4)$$

Here, $i$ is the delay time step of the feedback force and $\xi_n^x$ is the random Gaussian position fluctuations with mean 0 and with $\langle \xi_m^x \xi_n^x \rangle = 2D t_u \delta_{mn}$ where $\delta_{mn}$ is the Kronecker delta function. By squaring and averaging Eq. (4) [25], we derived the following relation between the variance $\langle x_n^2 \rangle$ and the virtual stiffness $k_v$,

$$\sigma^2 \equiv \langle x_n^2 \rangle = \frac{2Dt_u\left\{1+\beta\left[-\left(1-\frac{k_v}{k_{ot}}\right)\right]\right\}}{(2\beta-\beta^2)+(2\beta-2\beta^2+\beta^3)\left[-\left(1-\frac{k_v}{k_{ot}}\right)\right]-\beta^2\left[-\left(1-\frac{k_v}{k_{ot}}\right)\right]^2-\beta^3\left[-\left(1-\frac{k_v}{k_{ot}}\right)\right]^3} \quad (5)$$

where $\beta = t_u/\tau_R$ is the ratio of the updating time to the characteristic relaxation time $\tau_R = \gamma/k_{ot}$ for the particle in the OT. Fig. 3(e) and its inset show the plot of $\sigma^2$ and $k_{\text{eff}}$ as a function of $k_v$, respectively. The solid curve corresponding to Eq. (5) with $t_u$=10 μs agrees well with the experimental data. We found that initially $k_{\text{eff}}$ increases linearly with $k_v$, later deviates from the linear behavior and becomes maximum at $k_v = 200$ pN/μm. Accordingly, $k_{\text{eff}}$ is equal to $k_v$ only in the linear region, which is ~2.3 times larger than $k_{ot}$ for the current experimental conditions, providing an upper limit on $k_v$ for which the particle dynamics in virtual harmonic potential are close to the real harmonic potential. This observation agrees with the above presented power spectrum analysis. The linear region can be extended further by decreasing $k_{ot}$ and $t_u$ [26].

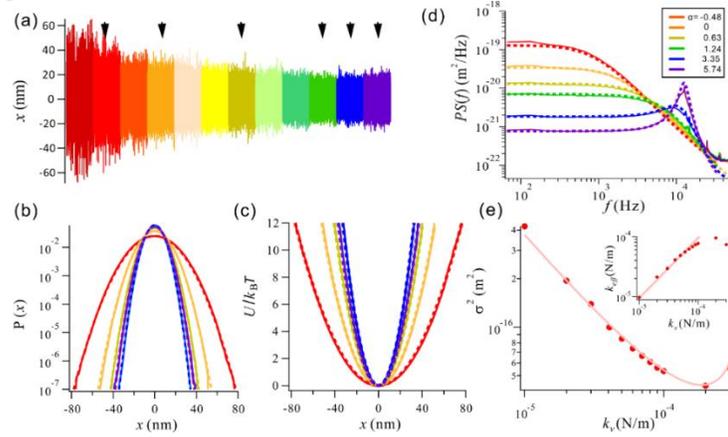

Fig. 3. (a) Trajectories of the particle in virtual harmonic potential for twelve different feedback gains $\alpha$. (b) Probability distributions of the particle position and (c) the corresponding harmonic potentials obtained by using Boltzmann distribution for six trajectories pointed by black arrows in panel (a). (d) Plot of power spectral density from same six trajectories in the panel (a). (e) Plot of the position variance of the particle as a function of $k_v$. The solid line is the theoretical prediction of Eq. (5). Inset is the plot of $k_{eff}$ as a function of $k_v$ and the solid line is the guideline of slope 1.

## 3. The virtual double-well potential

Next, we show that our feedback control scheme can generate a virtual double-well potential. The mathematical form of the double-well potential is given by

$$U_{dw}(x) = 4E_b\left[-\frac{1}{2}\left(\frac{x}{x_m}\right)^2 + \frac{1}{4}\left(\frac{x}{x_m}\right)^4 - A\frac{x}{x_m}\right], \quad (6)$$

where $\pm x_m$ is the position of the potential minima from the local maximum, $E_b$ is the barrier height separating two minima, and $A$ is the tilt amplitude of either well. The shift of the potential center corresponding to the feedback force $f_{dw} = -\partial_x U_{dw}(x)$ is given by $x_L(t) = 2x(t-t_d) - [x(t-t_d)]^3/x_m^2 + Ax_m$. Fig. 4(a) shows the typical trajectories of the particle in the virtual symmetric double-well potential ($A$= 0) recorded for three different barrier heights $E_b/k_BT = 2$ (red), 3 (green), and 4 (blue), respectively, with $x_m = 50$ nm. The probability distributions of the particle trajectories depicted in Fig. 4(b) have two nearly symmetric peaks located at two potential minima of $\pm 50$ nm. The measured probability distributions fit well to the Boltzmann distribution $P(x) \propto \exp(-U_{dw}(x)/k_B T)$. Fig. 4(c) shows the plots of the corresponding virtual double-well potentials fitting well to Eq. (6) (for the fit values, see Table 1). We also measured the average dwell time $\tau_D$ of the particle in each well as shown in Table 1. Our measured dwell time agrees well with the mean residence time predicted by the Kramers theory, $\tau_K = (\pi x_m^2)/(2\sqrt{2}\, E_b D) \exp(E_b/k_B T)$ [27]. Fig. 4(d) shows the plot of the virtual asymmetric double-well potentials tilted to the right (red) and to the left (blue) obtained from the probability distribution shown in the inset. The solid curves fit to the experimental data using Eq. (6) with $A < 0$ corresponding to the potential tilted to the right and vice versa. This means that the OFT can generate the mathematical double-well potential whose barrier height and well depths can be modulated in a controlled manner.

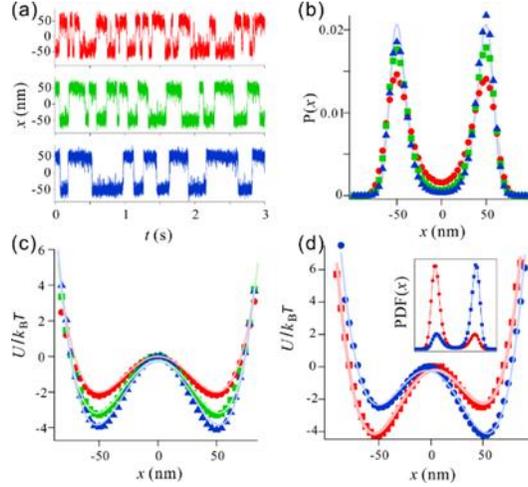

Fig. 4. Realization of a virtual double-well potential via optical feedback trap. (a) Trajectories of the particle in symmetric virtual double-well potential for $x_m = 50$ nm and $E_b/k_B T =$ 2 (red), 3 (green) and 4 (blue). (b) Probability distributions obtained from the particle trajectories and (c) the corresponding double-well potentials. (d) Double-well potentials tilted to the right (red) and left (blue) obtained from the probability distributions depicted in the inset.

| $(x_m)_{\text{fitted}}$ (nm) | $(E_b/k_B T)_{\text{fitted}}$ | $\tau_D$ (ms) | $\tau_k$ (ms) |
|---|---|---|---|
| $51 \pm 0.10$ | $2.03 \pm 0.03$ | $33 \pm 1$ | 43 |
| $52 \pm 0.06$ | $3.14 \pm 0.03$ | $79 \pm 3$ | 78 |
| $51 \pm 0.06$ | $3.94 \pm 0.03$ | $151 \pm 6$ | 159 |

Table 1 Parameters characterizing the virtual symmetric double-well potential. The errors refer to the standard error.

### 4. Nonequilibrium fluctuations in time-varying harmonic potential

As an application of the OFT, we study the nonequilibrium fluctuations of a Brownian particle in virtual harmonic potential whose stiffness is varied in time as shown in Fig. 5(a). Here, the virtual stiffness $k_v$ is varied linearly from the initial value $k_i$ to the final value $k_f$ during the driving time $\tau$. Prior study controlled the trap stiffness by adjusting the laser intensity [28]. This may change the bath temperature due to laser heating, and hence the particle dynamics may be affected. In addition, it could not modulate the trap stiffness at the faster rate. In comparison to the previous study, we change the trap stiffness in the much wider range of the driving time $\tau$ without changing the laser intensity.

The work performed on the particle when $k_v$ is varied from $k_i$ to $k_f$ is given by [29]

$$W = \int_{k_i}^{k_f} dk_v \frac{dU}{dk_v} = \frac{1}{2}\int_{k_i}^{k_f} dk_v x^2. \qquad (7)$$

If $\tau$ is much larger than the characteristic relaxation time $\tau_R$, the system is in the quasistatic state. Then, following the equipartition theorem, Eq. (7) can be written for $\tau > \tau_R$ as

$$\frac{W}{k_B T} = \frac{1}{2}\ln\left(\frac{k_f}{k_i}\right) \qquad (8)$$

which is equal to the free energy difference $\Delta F/k_B T$ between the final and the initial states. For $\tau < \tau_R$, the system is out of equilibrium, and $W/k_B T$ is no longer a fixed quantity, but rather has a distribution. Fig. 5(b) shows the average work performed on the system obtained by changing the virtual stiffness linearly from $k_i = 10$ pN/μm to $k_f = 40$ pN/μm (vice versa) for 13 different $\tau$ ranging from 50 μs to 1 s during the forward (reverse) process. The error bar represents the standard deviation. We found that, for $\tau \approx 10$ ms and greater, the average forward work (red circles) and the reverse work (blue squares) approach to the limit of $\Delta F/k_B T = \ln 2$ and $-\ln 2$, respectively, set by Eq. (8). For $\tau < 10$ ms, the average work deviates from the limit of $\Delta F/k_B T$ which explains the irreversibility of the nonequilibrium process. Our result agrees with the thermodynamic second law for isothermal process, $\langle W \rangle \geq \Delta F$. The solid curves are fit to $|\langle W \rangle|/k_B T = \Delta F/k_B T + B/\tau$ where $B$ is a constant and $\Delta F/k_B T$ is equal to $\ln 2$ at the given stiffness difference, which is a universal feature for the optimal driving scheme [30].

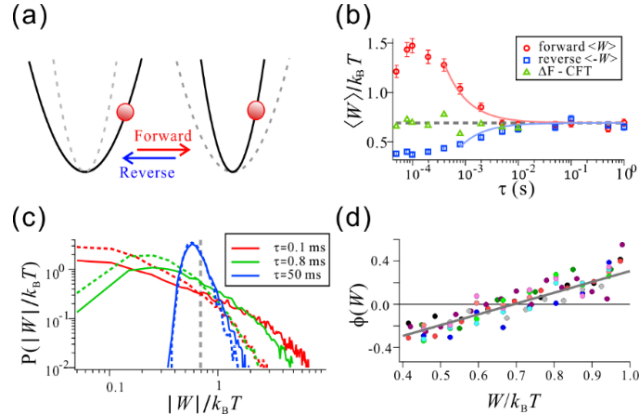

Fig. 5. (a) The scheme of the protocol showing the virtual stiffness of the virtual harmonic potential changes linearly from initial stiffness $k_i = 10$ pN/μm to $k_f = 40$ pN/μm (forward process) and vice versa (reverse process). (b) Plot of mean work (red circles: forward, blue squares: reverse) as a function of the driving time $\tau$. The solid curves are fit to $\langle W \rangle/k_B T = \Delta F/k_B T + B/\tau$ and the dashed horizontal line represents $\Delta F/k_B T = \ln 2$. The green triangles are the free energy difference $\Delta F/k_B T$ determined from (d). (c) The probability distribution for work during the forward (solid curves) and reverse (dashed curves) processes for $\tau$=0.1 (red), 0.6 (green), and 50 ms (blue). The dashed vertical line corresponds to $\Delta F/k_B T$. (d) Plot showing the validation of the Crook's fluctuation therorem for different driving times ranging from 50 μs to 50 ms (colored circles). The gray solid line is the plot of Eq. (9).

As a further check, we show that the above protocol of driving the system initially in thermal equilibrium to nonequilibrium state by changing the stiffness monotonically with time satisfies the Crooks fluctuation theorem [22]

$$\frac{P_F(W)}{P_R(-W)} = \exp\left(\frac{W - \Delta F}{k_B T}\right) \quad (9)$$

where $P_F(W)$ and $P_R(-W)$ are the forward and the reverse probability distributions for $W$, respectively. This equation explains the relation between the work performed on the system during the nonequilibrium process and the free energy difference. Fig. 5(c) shows the experimentally obtained work probability distribution during the forward ($W$, solid curves) and reverse ($-W$, dashed curves) processes obtained for $\tau = 0.1$ (red), 0.8 (green), and 50 ms (blue), respectively. The vertical gray dash line represents $\Delta F/k_B T$. The measured work distributions are strongly non-Gaussian in shape and are more skewed to the right of $\Delta F/k_B T$. For $\tau \gtrsim 10$ ms, $P_F(W)$ and $P_R(-W)$ nearly overlap; however, for shorter $\tau$, $P_F(W)$ is broader than $P_R(-W)$, intersecting each other at $\Delta F/k_B T$. Fig. 5(d) shows the plot of $\phi(W) \equiv \ln[P_F(W)/P_R(-W)]$ as a function of $W/k_B T$ agrees well with the theoretical prediction (gray solid line) of Crooks fluctuation theorem in Eq. (9). We also determined the free energy differences $\Delta F/k_B T$ for different $\tau$ by using Eq. (9), which shows the good agreement with $\Delta F/k_B T$ for the quasistatic process shown as green triangles in Fig. 5(b). The shortest updating time $t_u$ in previous study [28] was 1 ms (~1/10 of their quasistatic limit), which is 100 times larger than $t_u$=10 $\mu$s (1/1000 of quasistatic limit) in this study. Therefore, in comparison to the previous study which could test the validity of the Crooks fluctuation theorem only a little off the equilibrium state, current study demonstrated its validity in highly nonequilibrium regimes as well.

## 5. Conclusions

In conclusion, we developed a simple but effective technique based on the optical feedback force that can generate spatio-temporal potential with an arbitrary desired shape. We tested this technique by studying the dynamics of a Brownian particle in virtual harmonic potential whose stiffness can be modulated in a controlled manner by keeping the laser intensity constant, and confirmed that the dynamics of the particle in the virtual harmonic potential is very close to that of a genuine continuous potential. We also demonstrated the virtual double-well potential whose barrier height and well depths are controlled to the desired shape. The average dwell time of the particle in symmetric virtual double-well potential agrees with the Kramers' prediction. As an application of this technique, we tested the validity of Crooks fluctuation theorem by studying the nonequilibrium fluctuation dynamics of a particle in a harmonic potential whose stiffness is modulated with time ranging from near equilibrium process to highly nonequilibrium process.

Since the OFT can generate the time-varying potential of any desired shape by manipulating an optically trapped particle with a high precision position detection and ultrafast feedback control, it will be a powerful experimental tool to study various phenomena in the fields of nonequilibrium and information thermodynamics. For example, the OFT can study the thermodynamics of resetting in the nonharmonic potential that is difficult to create with other techniques [31], and the realization of Feynman ratchet which requires a spatial series of continuous but asymmetric potentials to be systematically manipulated [32,33]. Since the OFT can increase the effective stiffness of the harmonic potential, combining it with fluorescent microscopy will allow trapping of the submicron sized fluorescent particle at considerably lower laser power.

**Funding.** The MoST of Taiwan under the Grant No. 105-2112-M-008-026-MY3 (Y. J.) and The Korean government under the Grant No. IBS-R020-D1 (H. K. P.).

**Acknowledgment**. We would like to thank P. Y. Lai, YungFu Chen and Jin Tae Park for helpful discussion.